\documentclass[10pt,aps,prb,longbibliography,notitlepage]{revtex4-1}

\usepackage[latin1]{inputenc}
\usepackage[english]{babel}
\usepackage{amsmath,amssymb,graphicx}
\usepackage[letterpaper,bottom=1.5in]{geometry}

\newcommand{\res}[2]{\underset{#1}{Res}\left(#2\right)}
\renewcommand{\Re}{\mathrm{Re}\,}
\renewcommand{\Im}{\mathrm{Im}\,}

\begin{document} %%%%%%%%%%%%%%%%%%%%%%%%%%%%%%%%%%%%%%%%%%%%%%%%%%%%%%%%%%%%%%

\title{On the Time-Dependent Analysis of Gamow Decay}
\author{Detlef D\"urr}
 \affiliation{Mathematisches Institut, Ludwig-Maximilians-Universit\"at M\"unchen,\\ Theresienstr. 39, D-80333 M\"unchen, Germany}
\author{Robert Grummt}
 \email{grummt@math.lmu.de}
 \affiliation{Mathematisches Institut, Ludwig-Maximilians-Universit\"at M\"unchen,\\ Theresienstr. 39, D-80333 M\"unchen, Germany}
\author{Martin Kolb}
 \affiliation{Department of Statistics, University of Oxford,\\ 1 South Parks Road, Oxford OX1 3TG, United Kingdom}
\date{\today}

\begin{abstract}
  Gamow's explanation of the exponential decay law uses complex ``eigenvalues''
  and exponentially growing ``eigenfunctions''. This raises the question, how
  Gamow's description fits into the quantum mechanical description of nature,
  which is based on real eigenvalues and square integrable wave functions.
  Observing that the time evolution of any wave function is given by its expansion
  in generalized eigenfunctions, we shall answer this question in the most
  straightforward manner, which at the same time is accessible to graduate
  students and specialists. Moreover the presentation can well be used in physics
  lectures to students.
\end{abstract}
\maketitle

\section{Introduction}\label{sec:Intro}

In 1928 George Gamow considered the exponential decay of unstable atomic
nuclei.\cite{Gamow} His theoretical description was based on solutions of
the stationary Schr\"odinger equation
\begin{equation}\label{eq:StatSchr}
    H\psi=k^2\psi,\;\text{where}\quad H=-\frac{d^2}{dx^2}+V(x)
\end{equation}
and $V$ is a compactly supported potential.\footnote{We use units in which $m=1/2$ and $\hbar=1$}
Gamow's key idea was to describe decay with eigenfunctions $G(x)$ that, asymptotically,
behave like purely outgoing waves in the sense that
\begin{equation}\label{eq:GamowBC}
  \lim_{x\to\pm\infty}\big(G(x)-e^{\pm izx}\big)=0.
\end{equation}
This idea lead him to ``eigenfunctions'' with complex ``eigenvalues'' $z^2=E-i\Gamma$ ($E,\Gamma>0$) rather than real ones, i.e. $HG(x)=(E-i\Gamma)G(x).$ Assuming that the Gamow function $G(x)$ evolves according to the time dependent Schr\"odinger equation
\begin{equation}
  i\frac{\partial}{\partial t}G(t,x)= HG(t,x)=(E-i\Gamma)G(t,x),
\end{equation}
it decays exponentially in time $G(t,x)=e^{-iEt-\Gamma t}G(x).$

However, Gamow's description does not immediately connect with Quantum Mechanics. While Eq.~\eqref{eq:StatSchr} appears there, too, in Quantum Mechanics eigenvalues are real and wave functions are square integrable. Gamow functions, on the other hand, belong to complex eigenvalues and are not square integrable. In fact, it is readily seen from their purely outgoing behavior (Eq.~\eqref{eq:GamowBC}) and $z=\sqrt{E-i\Gamma}$ having negative imaginary part, that Gamow functions have exponentially growing tails. Such a function is not square integrable. So the question is: How does Gamow's description of exponential decay connect with Quantum Mechanics?

There are numerous mathematical articles concerned with this question, e.g. Refs.~\onlinecite{Costin,Lavine,Skibsted,Skibsted2}. From the articles it is, unfortunately, often not so easy to extract the clear and straightforward answer to that question. It is this: A Gamow function $G(x)$ is approximately a quantum mechanical generalized eigenfunction (i.e. scattering state). Since generalized eigenfunctions govern the time evolution of square integrable wave functions which are orthogonal to all bound states, there are special initial wave functions, namely those which are approximated by a Gamow function and which  therefore approximately undergo exponential decay in time.

Of course, this answer needs a bit of elaboration. We need to qualify the various 
``approximations'':  First, generalized eigenfunctions do not
have exponentially growing tails. Therefore, Gamow functions
approximate generalized eigenfunctions only locally, e.g. on the support of
the potential. The physical wave function, which undergoes approximate
exponential decay must be square integrable and therefore can only be locally
given by the Gamow function, too. Finally, approximate exponential decay in time
means that neither for very small nor for very large times exponential decay
holds. It only holds on an intermediate time regime.\footnote{A square
integrable wave function can not decay exponentially for small times because of
the unitarity of the time evolution operator $e^{-iHt}.$ Using the unitarity, we
can conclude for the survival probability $P_\psi(t)=|{\langle\psi,e^{-iHt}\psi\rangle}|^2$
that $P_\psi(t)\leq P_\psi(0).$ Since the survival probability is differentiable
and symmetric $P_\psi(-t)=P_\psi(t),$ this shows that $\frac{d}{dt}P_\psi(0)=0.$
Hence, exponential decay is impossible for very small times. In order to see
that it is impossible for very large times, too, we can use the fact that the
integrand in Eq.~\eqref{eq:TimeEvol} oscillates rapidly for large $t\gg1.$ This
leads to cancellations except when $k=0.$ Hence, Eq.~\eqref{eq:TimeEvol} can be
approximated by
\begin{equation*}
  e^{-iHt}\psi(x)\approx\big(\hat\psi_+(0)u_+(0,x)+\hat\psi_-(0)u_-(0,x)\big)t^{-1/2}\int_0^\infty e^{-i\kappa^2}d\kappa\sim t^{-1/2}
\end{equation*}
for $t\gg1,$ which shows that the wave function does not decay exponentially for
very large times either.}

Except for Ref.~\onlinecite{Roderich}, the pedagogical accounts on Gamow's
description of exponential decay we are aware of, usually only sketch its
connection to the quantum mechanical description based on square integrable wave
functions~\cite{Boehm,Cavalcanti,delaMadrid,Fuda,Holstein}. The purpose of our note
is to explain this connection in more detail. Compared to
Ref.~\onlinecite{Roderich}, which is a fairly complete discussion for a
particular potential, we will stress the general principles underlying the
connection between Gamow's description and the quantum mechanical description in a
way which seems the most straightforward one, and which will be accessible to
interested graduate students as well as specialists. The method presented here
can well be taken as starting point for further results concerning for example higher
dimensions. The presentation is also useful for teaching decay phenomena in
physics courses.

In the next Sec.~\ref{sec:Gamow} we give a heuristic explanation for Gamow
decay, which will subsequently be rigorously discussed. At the end of
Sec.~\ref{sec:Gamow} we will give an illuminating example for our arguments.

\section{Gamow functions and the Time Evolution of Square Integrable Wave Functions}
\label{sec:Gamow}

\begin{figure}[!ht]
  \centering
  \includegraphics[scale=.5]{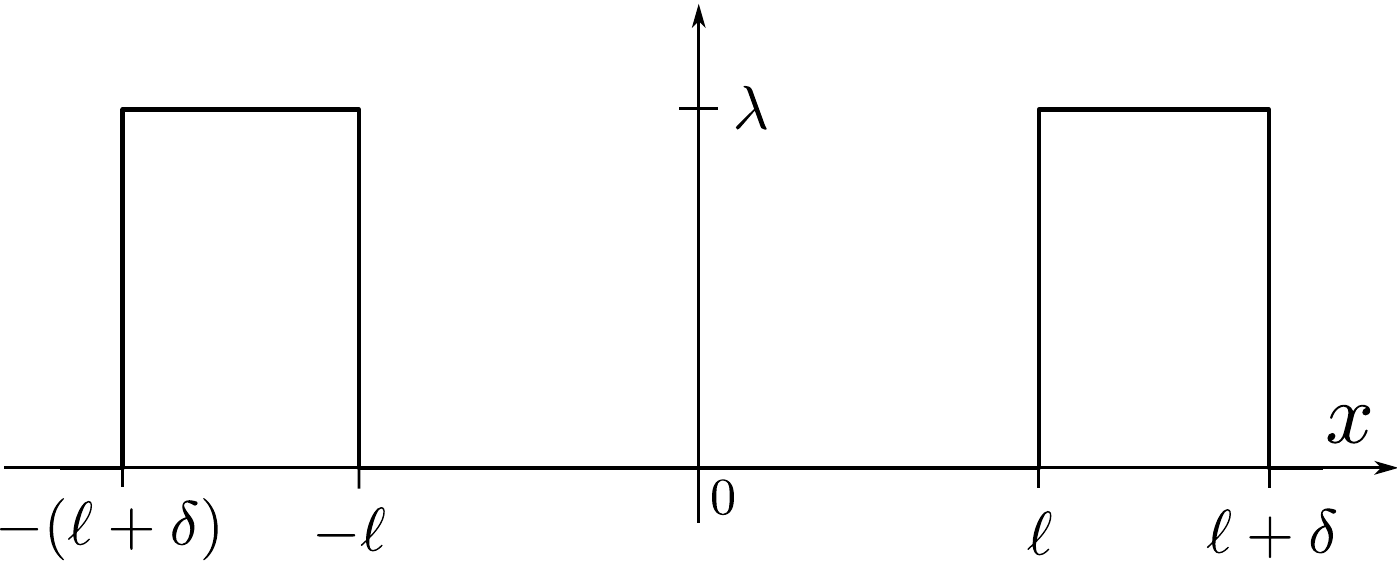}
  \caption{Plot of the double well potential.}
  \label{fig:Potential}
\end{figure}

Gamow had the right intuition, ``eigenfunctions'' corresponding to complex ``eigenvalues'' do give rise to long lived square integrable states, which decay exponentially in time. However, their presence becomes only apparent in special physical situations. The prototype of a potential that creates such a situation is the double well potential (see Fig.~\ref{fig:Potential}). Wave functions initially localized inside the double well are long lived if the wells are high, because at potential steps they are partially transmitted and partially reflected; if the steps are high, reflection outweighs transmission. At each time of transmission, it is natural to view the transmitted portion as being proportional to what is left inside the double well and thus exponential decay appears naturally.

This point of view shows that exponential decay is not at all a ``tunneling'' phenomenon, as it is often intuitively assumed. For a metastable state to occur, it suffices that a potential has steps at which a wave is partially reflected. The rectangular potential well $\lambda\mathbf 1_\ell$ is thus another example, which allows for unstable but long lived states that decay exponentially in time. Here $\mathbf 1_\ell(x)$ denotes the indicator function on $[-\ell,\ell];$ $\mathbf 1_\ell(x)$ equals one on~$[-\ell,\ell]$ and zero otherwise. If $\lambda$ is large, states initially localized on top of the rectangular potential well will be metastable and will decay exponentially in time.\cite{Diplom,Roderich} Gamow decay applies here as well and in Ref.~\onlinecite{Diplom} the decay is analyzed along the lines presented here. The picture of a wave being partially transmitted and partially reflected is somewhat hidden in Gamow's ansatz, but becomes more apparent when we relate the true quantum mechanical time evolution of the meta stable state to the Gamow function.

If the double well (Fig.~\ref{fig:Potential}) allows for a Gamow function $G,$ then the truncated version of it, namely  $\mathbf 1_\ell G$
yields a long lived square integrable initial wave function. The question we address is: Does $\mathbf 1_\ell G$ decay exponentially
in time? In this section, we will explain that, on intermediate time scales, it actually does. At first we will consider general one dimensional potentials, but at the end (Section~\ref{sec:Example}) we will return to the concrete example of the double well potential. Throughout this note we will only consider potentials $V,$ which have compact support contained in $[-L,L]$ (note that $L\neq\ell$) and since exponential decay is a genuine scattering phenomenon, we also assume that the potentials have no bound states.

\subsection{Heuristic Argument}\label{sec:Heuristics}

Assume that $V$ allows for a Gamow function $G.$ We will give a two step
argument, which shows that $\mathbf 1_LG$ decays exponentially in time when
evolved according to the time dependent Schr\"odinger equation
$i\partial_t\psi=H\psi.$ The general solution of the time dependent Schr\"odinger
equation is given by $e^{-iHt}\psi.$ In the first step we will establish a
generic connection between the time evolution of any square integrable wave
function and the Gamow function $G.$ In the second step we will use this
connection to show that $e^{-iHt}\mathbf 1_LG$ decays exponentially in time.

So, how does $e^{-iHt}\mathbf 1_LG$ evolve in time? To find an answer, we need
a method that makes the time evolution palpable. For this purpose, we will use
the method of expansions in generalized eigenfunctions $\varphi(k,x),$ which
applied to an arbitrary square integrable wave function $\psi$ yields
\begin{align}
  \label{eq:EigenfctExpansion}
  &\psi(x)=\int^\infty_{-\infty}\hat\psi(k)\varphi(k,x)\,dk\quad\text{with}\\
  &\hat\psi(k)=\int^\infty_{-\infty}\psi(x)\overline{\varphi(k,x)}\,dx.
\end{align}
These generalized eigenfunctions are bounded, but not square integrable
solutions to the stationary Schr\"odinger equation
\begin{equation}\label{eq:Schroedinger}
 H\varphi(k,x)=\bigg(-\frac{d^2}{dx^2}+V(x)\bigg)\varphi(k,x)=k^2\,\varphi(k,x).
\end{equation}
An expansion in terms of $\varphi$ diagonalizes $H$ in a completely analogous
way as the Fourier transform diagonalizes $-\tfrac{d^2}{dx^2}.$ The time evolved
$\psi$ can thereby be expressed in a very concrete analytical way
\begin{equation}\label{eq:Expansion}
  e^{-iHt}\psi(x)=\int^\infty_{-\infty}\hat\psi(k)\varphi(k,x)e^{-ik^2t}\,dk.
\end{equation}

Why should the time evolution expressed in terms of an expansion in generalized
eigenfunctions~\eqref{eq:Expansion} be related in any way to the Gamow function
$G?$ Because both, the generalized eigenfunctions $\varphi$ as well as the Gamow
function $G,$ solve the stationary Schr\"odinger equation~\eqref{eq:Schroedinger};
the Gamow function for complex $k^2=E-i\Gamma$ with $E,\Gamma>0$ and the
generalized eigenfunctions for real $k^2\geq0.$ This suggests that in some sense
$\varphi\approx G,$ when the complex ``eigenvalue'' is close to the real axis
($\Gamma\ll1$). According to Eq.~\eqref{eq:Schroedinger} generalized eigenfunctions
behave like plane waves in regions where the potential is zero. Therefore, combining
plane wave behavior and "near Gamow function behavior", we make the ansatz
\begin{equation}\label{eq:Ansatz}
  \varphi(k,x)\approx\eta(k)\mathbf 1_LG(x)+e^{ikx}.
\end{equation}
We need to determine $\eta.$ Plugging Eq.~\eqref{eq:Ansatz} into
Eq.~\eqref{eq:Schroedinger}, we find
\begin{equation}
  \bigg(-\frac{d^2}{dx^2}+V\bigg)(\eta(k)\mathbf 1_LG+e^{ikx})\approx k^2\,(\eta(k)\mathbf 1_LG+e^{ikx}).
\end{equation}
Now, $H\mathbf 1_LG\approx z^2\mathbf 1_LG$ and
$-\frac{d^2}{dx^2}e^{ikx}=k^2e^{ikx},$ so we can rearrange the above
equation, putting $f(k,x)=V(x)e^{ikx},$ such that
\begin{equation}
  \eta(k)\mathbf 1_LG(x)\approx\frac{f(k,x)}{k^2-z^2}.
\end{equation}
Integrating both sides with respect to $x,$ entails that
\begin{equation}\label{eq:Eta}
  \eta(k)\approx\frac{\tilde f(k)}{k^2-(E-i\Gamma)},
\end{equation}
where $\tilde f$ is some analytic function and $z^2=E-i\Gamma.$ We find that the complex ``eigenvalue'' $E-i\Gamma$ causes the generalized eigenfunctions $\varphi(k,x)$ to have a pole, when continued to the complex $k$-plane. The Gamow function is the corresponding residue. This was the first step of our heuristic argument.

In the second step, we will use Eq.~\eqref{eq:Eta} in Eq. \eqref{eq:Ansatz} to show
that $e^{-iHt}\mathbf 1_LG$ decays exponentially in time. We only need to calculate the integral in the eigenfunction
expansion~\eqref{eq:Expansion}. The heart of this calculation lies in the fact that
the first summand on the right hand side of~\eqref{eq:Ansatz} dominates when
$\Gamma\ll1,$ because $|\eta(k)|$ is much larger than $|e^{ikx}|$ for $k\approx\Re z.$ Therefore,
\begin{align}
\varphi(k,x)\approx\eta(k)\mathbf 1_LG(x)
\end{align}
and hence
\begin{align}
  \widehat{\mathbf 1_LG}(k)&\approx\int_{-\infty}^\infty\mathbf
  1_LG(x)\overline\eta(k)\mathbf 1_LG(x)\,dx\approx c\,\overline\eta(k),\nonumber\\
  e^{-iHt}\mathbf 1_LG(x)&\approx c\int^\infty_{-\infty}\overline\eta(k)\eta(k)\mathbf 1_LG(x)e^{-ik^2t}\,dk\nonumber\\
  &= c\,\mathbf 1_LG(x)\int^\infty_{-\infty}\frac{|\tilde f(k)|^2}{|k^2-(E-i\Gamma)|^2}e^{-ik^2t}\,dk.\label{eq:Residue}
\end{align}
\begin{figure}[!t]
  \centering
  \includegraphics[scale=.75]{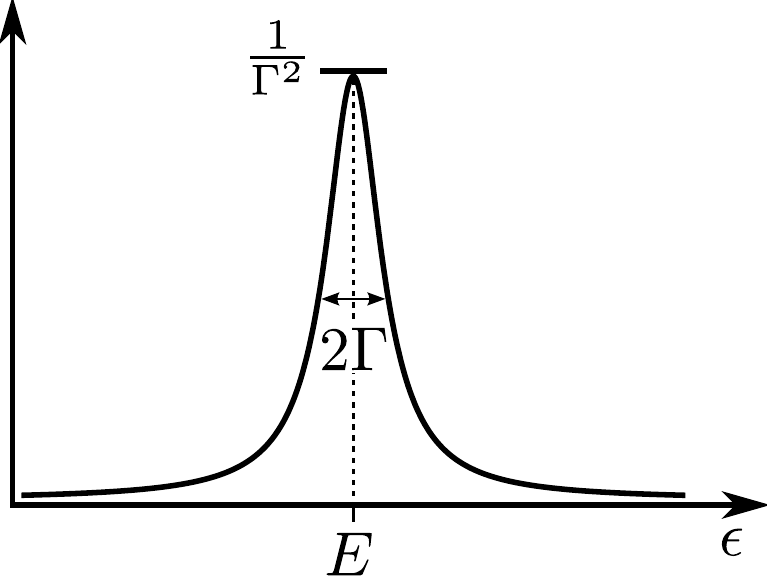}
  \caption{Plot of the Breit-Wigner function $\frac{1}{(\epsilon-E)^2+\Gamma^2}.$}
  \label{fig:BreitWigner}
\end{figure}
To solve the integral notice that it is essentially the Fourier transformation of the Breit-Wigner function $1/\big[(k^2-E)^2+\Gamma^2\big].$ They differ only by the appearance of an additional function $|\tilde f(k)|^2$ and the fact that in Eq.~\eqref{eq:Residue} we integrate over $k$ instead of $k^2.$ Therefore, we change the integration variable
\begin{align}
  e^{-iHt}\mathbf 1_LG(x)&\approx c\,\mathbf 1_LG(x)\int^\infty_{-\infty}\frac{|\tilde f(k)|^2}{|k^2-(E-i\Gamma)|^2}e^{-ik^2t}\,dk\nonumber\\
  &=c\,\mathbf 1_LG(x)\bigg[\int^\infty_{0}\frac{|\tilde f(k)|^2}{|k^2-(E-i\Gamma)|^2}e^{-ik^2t}\,dk+\int^0_{-\infty}\frac{|\tilde f(k)|^2}{|k^2-(E-i\Gamma)|^2}e^{-ik^2t}\,dk\bigg]\nonumber\\
  &=c\,\mathbf 1_LG(x)\bigg[\int^\infty_{0}\frac{|\tilde f(+\sqrt\epsilon)|^2}{(\epsilon-E)^2+\Gamma^2}e^{-i\epsilon t}\,\frac{d\epsilon}{2\sqrt\epsilon}+\int^\infty_0\frac{|\tilde f(-\sqrt\epsilon)|^2}{(\epsilon-E)^2+\Gamma^2}e^{-i\epsilon t}\,\frac{d\epsilon}{2\sqrt\epsilon}\bigg].
  \label{eq:Residue2}
\end{align}
We substituted $k=\sqrt\epsilon$ in the first integral and $k=-\sqrt\epsilon$ in the second. Due to the fact that the Breit-Wigner function is strongly peaked at $\epsilon=E$ if $\Gamma\ll1$ (see Fig.~\ref{fig:BreitWigner}), the integrand in Eq.~\eqref{eq:Residue2} is localized about $E>0.$ Hence, we can replace $\tilde f(\pm\sqrt\epsilon)$ and $1/\sqrt\epsilon$ by their respective values at $\epsilon=E,$ so that
\begin{align}
  e^{-iHt}\mathbf 1_LG(x)&\approx c\,\mathbf 1_LG(x)\bigg[\frac{|\tilde f(+\sqrt E)|^2}{2\sqrt E}+\frac{|\tilde f(-\sqrt E)|^2}{2\sqrt E}\bigg]
  \int^\infty_{0}\frac{1}{(\epsilon-E)^2+\Gamma^2}e^{-i\epsilon t}\,d\epsilon\nonumber\\
  &\approx c'\,\mathbf 1_LG(x)\,e^{-\Gamma t},\label{eq:MainResult}
\end{align}
where we have used that the Fourier transformation of the Breit-Wigner function is the exponential function. Thus, $e^{-iHt}\mathbf 1_LG$ decays exponentially in time whenever $\Gamma\ll1.$

\subsection{Towards Rigor}\label{sec:Rigor}

The crucial point in our heuristic argument is the form of the generalized eigenfunctions, i.e. the ansatz~\eqref{eq:Ansatz} together with the function $\eta$ from Eq.~\eqref{eq:Eta}, because the time evolution is determined by the way the generalized eigenfunctions look like. How can this be made precise?

First of all, the expansion in generalized eigenfunctions is defined by
\begin{align}
  \label{eq:InvFourier}
  &\psi(x)=\int_0^\infty(\hat\psi_+(k)u_+(k,x)+\hat\psi_-(k)u_-(k,x))\,dk\quad\text{with}\\
  &\hat\psi_\pm(k)=\int_{-\infty}^\infty\psi(x)\overline{u_\pm(k,x)}\,dx,\qquad k\in[0,\infty)\label{eq:Fourier}
\end{align}
and since it diagonalizes $H,$ we have
\begin{equation}\label{eq:TimeEvol}
  e^{-iHt}\psi(x)=\int_0^\infty(\hat\psi_+(k)u_+(k,x)+\hat\psi_-(k)u_-(k,x))\,e^{-ik^2t}\,dk
\end{equation}
for every square integrable wave function $\psi.$ Here $u_\pm(k,x)$ solve Eq.~\eqref{eq:Schroedinger}. Eq.~\eqref{eq:InvFourier} differs from Eq.~\eqref{eq:EigenfctExpansion} by the appearance of an additional summand and by integrating from $0$ to $\infty$ rather than from $-\infty$ to $+\infty.$ This form of Eq.~\eqref{eq:InvFourier} is due to Schr\"odinger's equation~\eqref{eq:Schroedinger} being an ordinary differential equation of second order. As such it has two linearly independent solutions for every $k^2\in\mathbb C.$ In particular, it has two linearly independent generalized eigenfunctions, $u_+(k,x)$ and $u_-(k,x),$ for every $k^2\geq0$ and both are needed to have a complete basis.\footnote{The two linearly independent generalized eigenfunctions form a complete basis only if $H$ has no bound states. In case $H$ has bound states, the bound states need to be added to the two linearly independent generalized eigenfunctions to get a complete basis.} For some potentials $u_+(-k,x)=u_-(k,x),$ e.g. when $V=0$ for which $u_+=e^{ikx}/\sqrt{2\pi}$ and $u_-=e^{-ikx}/\sqrt{2\pi}$. Then Eq.~\eqref{eq:InvFourier} can be rewritten in shorter form (like Eq.~\eqref{eq:EigenfctExpansion}) with an integral extending from $-\infty$ to $+\infty,$ but in general this is not possible. For more details on expansions in generalized eigenfunctions see Ref.~\onlinecite{[] [{, Theorem~8.4}]Weidmann2engl}.

A potential with finite range perturbs the free Hamiltonian merely on its range, so that the generalized eigenfunctions are essentially plane waves, distorted only on the range of the potential. Hence, we make the usual~\cite{[] [{, Chapter~2.6}]Griffiths} ansatz to obtain the precise generalized eigenfunctions:
\begin{align}
  &f_+(k,x)=
  \begin{cases}
	a_+(k)e^{ikx}+b_+(k)e^{-ikx}\quad&,x<-L\\
	c_+(k)e^{ikx}&,x>L
  \end{cases}\label{eq:f_+}\\
  &f_-(k,x)=
  \begin{cases}
	c_-(k)e^{-ikx}\;\,\quad\qquad\qquad\quad&,x<-L\\
	a_-(k)e^{-ikx}+b_-(k)e^{ikx}&,x>L
  \end{cases}\label{eq:f_-}
\end{align}
with $k\in[0,\infty).$ Physically, Eq.~\eqref{eq:f_+} is a wave incident from the left together with its transmitted and reflected part; analogously Eq.~\eqref{eq:f_-} is a wave incident from the right. Now, it would be natural to set $a_\pm=1,$ thereby normalizing the amplitude of the incident wave to one, because then $|b_\pm|^2$ and $|c_\pm|^2$ would be the reflection- and transmission coefficients, respectively. 

However, we choose a different route by setting $c_\pm=1.$ Then the functions $f_\pm, a_\pm$ and $b_\pm$ are analytic in~$k,$ because $f_+$ ($f_-$) is analytic for $x>L$ ($x<-L$) and therefore needs to be analytic in $k$ for all $x.$ Thus $f_\pm$ have no poles. The poles occur upon normalization. Let $u_\pm$ denote the normalized eigenfunctions, then they need to satisfy
\begin{equation}\label{eq:Normalization}
  \delta(k-k')=\int_{-\infty}^\infty u_\pm(k,x)\overline{u_\pm(k',x)}\,dx,
\end{equation}
which they do if
\begin{equation}\label{eq:GEF}
  u_\pm(k,x)=\frac{1}{\sqrt{2\pi}}\frac{f_\pm(k,x)}{a(k)},\qquad k\in[0,\infty).
\end{equation}
Here $a:=a_+=a_-.$ To see that $a_+$ is identical to $a_-,$ note that the Wronskian $W(f_+,f_-)=f_+f_-'-f_+'f_-$ is independent of $x.$\footnote{The Wronskian $W(f_+,f_-)$ is independent of $x,$ since $f_\pm$ are solutions of the Schr{\"o}dinger equation.} Thus, we can calculate the Wronskian once using the $f_\pm$ for $x>L$ and another time using the $f_\pm$ for $x<-L;$ the results must be
identical and this leads us to $a_+=a_-.$ Note furthermore that the transmission- and reflection coefficients are now given by
\begin{equation}
  T(k)=\frac{1}{|a_\pm(k)|^2}=\frac{1}{|a(k)|^2}\qquad\text{and}\qquad R_\pm(k)=\frac{|b_\pm(k)|^2}{|a(k)|^2},
\end{equation}
respectively. Due to the fact that we set $c_\pm=1,$ the analytic structure of the generalized eigenfunctions is now evident: $u_\pm$ have poles whenever $a=0,$ because setting $c_\pm=1$ ensures that $f_\pm$ and $a$ have no poles themselves.\footnote{The point $k=0$ often needs special attention in scattering theory. Thus, there might be situations in which $f_\pm$ and $a$ admit analytic extensions only to the punctured complex plane $\mathbb C\setminus\{0\}.$ However, this is no obstacle for our argument.}

We would like to argue now that Gamow functions are the residua of generalized eigenfunctions. From Eq.~\eqref{eq:GEF} we see that $u_\pm$ only have poles if $a=0.$ If $z$ denotes a root of $a(k),$ then the residue of $u_\pm$ at $z$ is (up to a constant) given by $f_\pm(z,x).$ To see that $f_\pm(z,x)$ satisfies the purely outgoing boundary condition that Gamow functions satisfy (Eq.~\eqref{eq:GamowBC}), observe that $a=0$ does not only cause the incident waves to vanish, but also $b_\pm=1.$ For, $W(f_+,f_-)=-2ik\,a(k).$ So whenever $a$ vanishes, the Wronskian vanishes, too. This is only possible if $f_+$ is a multiple of $f_-,$ which by a comparison of Eq.~\eqref{eq:f_+} with Eq.~\eqref{eq:f_-} implies that $b_\pm(z)=1.$ Thus, $f_\pm(z,x)$ satisfies the purely outgoing boundary condition~\eqref{eq:GamowBC}. Observing that $f_\pm(z,x)$ solves Eq.~\eqref{eq:Schroedinger}, too, we conclude that it is a Gamow function, so that
\begin{equation}\label{eq:GamowFunction}
  \res{k=z}{u_\pm(k,x)}=c\,f_\pm(z,x)=c\,G(x).
\end{equation}
Moreover, since
\begin{equation}
  HG(x)=Hf_\pm(z,x)=z^2f_\pm(z,x)=z^2G(x),
\end{equation}
$z^2$ is the corresponding eigenvalue. Eigenvalues corresponding to Gamow functions are usually called resonance. We will use this term not only for $z^2,$ but also for $z.$

Our ansatz~\eqref{eq:Ansatz} is now rigorously replaced by the Laurent series of $u_\pm(k,x)$ about the root $z$ of $a(k)$
\begin{equation}\label{eq:Laurent}
  u_\pm(k,x)=\eta(k)G(x)+a^\pm_0(x)+a^\pm_1(x)(k-z)+\dots,
\end{equation}
where $\eta(k)=c/(k-z).$ But notice the difference to our
ansatz~\eqref{eq:Ansatz}, where the Gamow function comes with the factor
$\mathbf 1_L,$ which truncates the exponential tails of $G.$ In the Laurent
Expansion~\eqref{eq:Laurent} this factor does not appear. Since $u_\pm$ is
bounded and $G$ increases exponentially, the remainder must compensate the
exponential tails of $G$ for large $x,$ so that the whole sum remains bounded.
The principal part of the Laurent Expansion therefore dominates only locally on bounded intervals. That
is sufficient for our purposes, because we are interested in the exponential
decay regime, where the majority of the wave function's mass lingers in
intervals around the range of the potential.

We wish to explain now under which conditions the principal part of the Laurent expansion governs the time evolution, so that exponential decay becomes dominant. For that purpose we consider the time evolution of $\psi=\mathbf 1_LG,$ where $G$ shall denote the Gamow function with eigenvalue $z^2=E-i\Gamma.$ Separating the principal part from the remainder of the Laurent expansion~\eqref{eq:Laurent}, we obtain for the eigenfunction expansion~\eqref{eq:TimeEvol}
\begin{align}
  \mathbf 1_Le^{-iHt}\mathbf 1_LG(x)=&\mathbf
  1_L\int_0^\infty\big(\widehat{\mathbf 1_LG}_+(k)+\widehat{\mathbf 1_LG}_-(k)\big)\eta(k)G(x)e^{-ik^2t}\,dk\label{eq:Main}\\
  &+\mathbf 1_L\int_0^\infty\big(\widehat{\mathbf 1_LG}_+(k)(u_+(k,x)-\eta(k)G(x))\label{eq:Error}\\
  &\quad\qquad\qquad+\widehat{\mathbf 1_LG}_-(k)(u_-(k,x)-\eta(k)G(x))\big)e^{-ik^2t}\,dk.\nonumber
\end{align}
Here we have inserted an additional factor $\mathbf 1_L$ for the reasons we have
just explained. We have chosen the indicator function on the range of the
potential only for simplicity. Regions growing with time would have been
possible, too.\cite{Roderich,Skibsted}

\begin{figure}[!th]
  \centering
  \includegraphics[scale=.75]{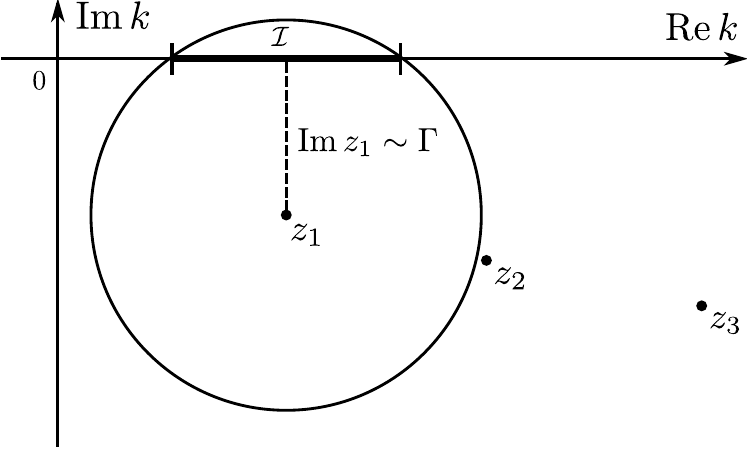}
  \caption{The disk on which the Laurent series for $u_\pm$ converges. Note that the figure also illustrates that higher order resonances have larger decay rates.}
  \label{fig:Laurent}
\end{figure}
The Gamow contribution~\eqref{eq:Main} dominates the time evolution whenever
$|\widehat{\mathbf 1_LG}_\pm(k)|$ is strongly peaked at $k=\Re z$ and if
$\Gamma\ll1,$ because then~\eqref{eq:Error} is negligible. We need both
conditions, since the Laurent expansion of $u_\pm$ is only valid for those $k$
for which the series converges. As $V$ might have more than one resonance, the
radius of convergence is in general finite, so that the Laurent expansion can
only be used on a bounded interval $\mathcal I\subset\mathbb R$ (see
Fig.~\ref{fig:Laurent}). The contribution from this interval $\mathcal I$
to~\eqref{eq:Error} can be neglected for small decay rates $\Gamma\ll1,$ since
then the resonance is close to the real axis, in which case the remainder of the
Laurent expansion is negligible compared to its principal part. The contribution
from $\mathcal I^c=[0,\infty)\setminus\mathcal I$ can be neglected when
$\widehat{\mathbf 1_LG}_\pm$ is strongly peaked at $k=\Re z,$ because then
$\widehat{\mathbf 1_LG}_\pm$ is small on $\mathcal I^c.$ That $\widehat{\mathbf
1_LG}_\pm$ actually is peaked at $k=\Re z$  comes from the fact that
$\widehat{\mathbf 1_LG}_\pm$ accumulates around $k=\Re z,$ because for these $k$
we have $u_\pm\approx G$ when $\Gamma\ll 1.$ Putting the contributions from
$\mathcal I$ and $\mathcal I^c$ together, we see that neglecting the
integral~\eqref{eq:Error} causes a small error whenever $\Gamma\ll1.$

We refer to our heuristic Section~\ref{sec:Heuristics}, to see now how the
exponential decay arises from the principal part of the Laurent series. There is
not much more to say on that.

\subsection{Example}\label{sec:Example}

Now we consider the double well potential (see Fig.~\ref{fig:Potential}). Also in this section we work with units in which $m=1/2$ and $\hbar=1$ to keep formulas short. However, to provide a feeling for dimensions we will give some of the results in SI units. In this regard, recall that we study the dynamics of $\alpha$-particles ($\text{He}^4_2$ nuclei). The mass $m$ that appears in Schr\"odinger's equation if it is written in SI units, therefore does not refer to electron mass, but the mass of a helium nucleus, which is $6.69\times10^{-27}$kg. The conversion between our units and SI units then follows the scheme: $x_{\mathrm{SI}}=a_0\,x,$ $k_{\mathrm{SI}}=k/a_0,$ $\lambda_{\mathrm{SI}}=|E_1|\,\lambda$ and $t_{\mathrm{SI}}=(\hbar/|E_1|)\,t,$ where $a_0=7.2\times10^{-15}$m and $E_1=-0.1$ MeV.

Due to the simplicity of the double well potential, the generalized eigenfunctions can be calculated explicitly. For brevity we only give
\begin{align}
  \label{eq:ExampleGEF}
  &u_+(k,x)=\frac{1}{\sqrt{2\pi}}\frac{1}{a(k)}
  \begin{cases}
	a(k)\,e^{ikx}+b_+(k)\,e^{-ikx}\quad &,x<-(\ell+\delta)\\
	a_1(k)\cos(kx)+i\,a_2(k)\sin(kx) &,|x|\leq\ell\\
	e^{ikx} &,x>\ell+\delta,
  \end{cases}
\end{align}
where $a(k)=a_1(k)a_2(k)$ (the full formula and the auxiliary functions are given in the Appendix). This example perfectly illustrates the structure of the generalized eigenfunctions that we have found in Section~\ref{sec:Rigor}. In particular, it reflects that $u_+$ has poles whenever the function $a(k)$ has roots. The decay rates (in SI units) of the corresponding metastable states are connected to the imaginary part of these roots by the formula
\begin{align}\label{eq:DecayRate}
  \Gamma_{\text{SI}}=\frac{|E_1|}{\hbar}\Gamma=\frac{|E_1|}{\hbar}\Im z^2,
\end{align}
where $z$ denotes a root of $a(k).$

Let us illustrate the decay rates of the double well potential using experimental data. Uranium~$234$ for example, has an experimentally measured decay rate of $\Gamma_{\text{exp}}=1.2936\times10^{-13}$ 1/s (see Ref.~\onlinecite[Table~1]{Duarte}). To model $\alpha$-decay of $\text{U}^{234}$ with the double well potential, we choose parameters which seem reasonable from what we know experimentally: For $\ell$ we choose the radius of the $\text{U}^{234}$ nucleus, i.e. $\ell=1$ ($\ell_{\text{SI}}=7.2\times10^{-15}$m). For $\delta$ we choose the value $2\ell$ ($\delta_{\text{SI}}=14.4\times10^{-15}$m). And $\lambda$ shall have the same order of magnitude as the Coulomb repulsion experienced by the $\alpha$-particle at radius $\ell$ ($V_{\text{Coulomb}}(\ell)=36$ MeV), but its particular value is fitted such that the theoretical decay rate is close to the experimental decay rate. The best value is $\lambda=436$ ($\lambda_{\text{SI}}=43.6$ MeV). We calculate the roots of $a(k)$ and thereby the decay rates of the double well potential numerically (we have used Mathematica\textsuperscript{\textregistered}). The root that describes the decay of $\text{U}^{234}$ best is
\begin{align}\label{eq:DecayRateEx}
  \hbar z_{\text{SI}}=1.0967\times10^{-19}-i\,8.5951\times10^{-55}\,(\text{m/s})
\end{align}
and the corresponding decay rate calculates to $\Gamma_{\text{SI}}=1.3361\times10^{-13}\tfrac{1}{\text{s}},$ which is very close to the experimental value.
\begin{figure}[!t]
  \centering
  \includegraphics[scale=.48]{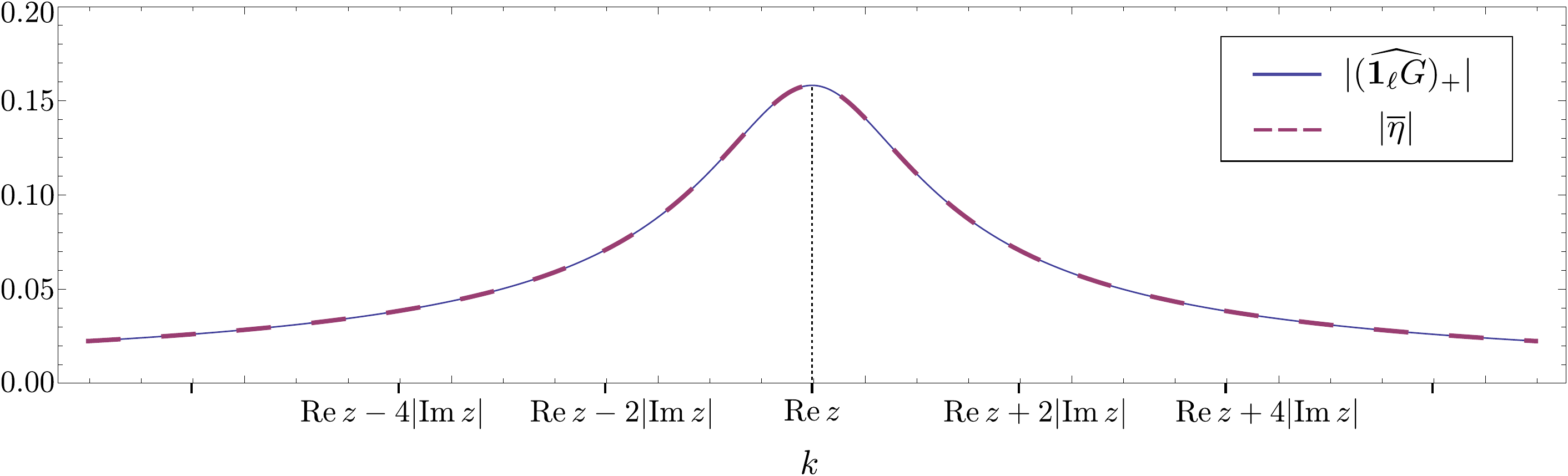}
  \caption{The modulus of the generalized Fourier transform of $\mathbf 1_\ell G$ (solid thin line) in comparison to $|\overline\eta|$ (dashed thick line) is shown. For the figure we have used the same parameter values as for the Uranium~$234$ example, which is discussed in the text (i.e. $\ell_{\text{SI}}=7.2\times10^{-15}\text{m}, \delta_{\text{SI}}=14.4\times10^{-15}\text{m}, \lambda_{\text{SI}}=43.6\;\text{MeV}$). Here $z$ denotes the root of $a(k)$ from Eq.~\eqref{eq:DecayRateEx}. We omitted the index SI for brevity.}
  \label{fig:PlotGamow}
\end{figure}

From the generalized eigenfunctions, we obtain the Gamow functions by
Eq.~\eqref{eq:GamowFunction} and the fact that $b_\pm=1$ if $a=0$ (see
Sec.~\ref{sec:Rigor}),
\begin{align}
 \label{eq:ExampleGamow}
  G(x)=c
  \begin{cases}
	e^{-izx}\quad &,x<-(\ell+\delta)\\
	a_1(z)\cos(zx)+i\,a_2(z)\sin(zx) &,|x|\leq \ell\\
	e^{izx} &,x>\ell+\delta,
  \end{cases}
\end{align}
where $z$ denotes a root of $a(z).$ It is immediately evident that $G$ satisfies the purely outgoing boundary condition~\eqref{eq:GamowBC}. Moreover, we see that the Gamow functions have exponentially increasing tails, confirming our considerations in the introduction (Sec.~\ref{sec:Intro}).

Using Eq.~\eqref{eq:ExampleGEF} for the generalized eigenfunction and Eq.~\eqref{eq:ExampleGamow} for the Gamow function, we can calculate the generalized Fourier transform $(\widehat{\mathbf 1_\ell G})_+$ via Eq.~\eqref{eq:Fourier}. This allows us to check that $(\widehat{\mathbf 1_\ell G})_+\approx\overline\eta,$ which we have used in Sec.~\ref{sec:Heuristics}. Due to the fact that $a=a_1a_2,$ either $a_1$ or $a_2$ vanishes when evaluated at a resonance. Suppose $z$ is a resonance for which $a_2$ vanishes, then
$\mathbf 1_\ell G(x)=c\,\mathbf 1_\ell\cos(zx)$ and
\begin{align}
  (\widehat{\mathbf 1_\ell G})_+(k)&=c\int_{-\ell}^\ell\cos(zx)\frac{\cos(kx)}{\overline a_2(k)}\,dx-i\,c\int_{-\ell}^\ell\cos(zx)\frac{\sin(kx)}{\overline a_1(k)}\,dx\\
  &=c\int_{-\ell}^\ell\cos(zx)\frac{\cos(kx)}{\overline a_2(k)}\,dx\\
  &=\frac{c}{\overline a_2(k)}\bigg(\frac{\sin((k-z)\ell)}{k-z}+\frac{\sin((k+z)\ell)}{k+z}\bigg),
\end{align}
where we have used that $\cos(zx)\sin(kx)$ is antisymmetric in the second step. In Fig.~\ref{fig:PlotGamow} we have plotted the modulus of $(\widehat{\mathbf 1_\ell G})_+$ in comparison to the modulus of $\overline\eta=c/(k-\overline z).$ The amazing agreement confirms that $(\widehat{\mathbf 1_\ell G})_+\approx\overline\eta.$

\section{Conclusion}

Although, Gamow's approach based on complex ``eigenvalues'' directly leads to
the exponential decay law, it involves exponentially growing ``eigenfunctions''.
Therefore, Gamow's approach to exponential decay can not be taken at face value.
However, Gamow's exponentially growing ``eigenfunctions'' are approximately
quantum mechanical generalized eigenfunctions of the Hamiltonian. Since
generalized eigenfunctions govern the time evolution of square integrable wave
functions, Gamow functions give rise to special (square integrable) initial wave
functions, which approximately undergo exponential decay in time. So, Gamow's
approach taken with a grain of salt yields in fact the explanation of the exponential
decay law.
\begin{acknowledgments}
  We should like to thank P. Pickl for many valuable discussions that shaped
  some of the ideas presented in this note. R.G. acknowledges financial
  support from Studienstiftung des deutschen Volkes as well as from the
  International Max-Planck-Research School of Advanced Photon Science.
\end{acknowledgments}

\begin{appendix}
\section*{Appendix: Generalized Eigenfunctions of Double Well Potential}

We now give the full formula for the generalized eigenfunctions of the double well potential
\begin{align}
  &u_+(k,x)=\frac{1}{\sqrt{2\pi}}\frac{1}{a(k)}
  \begin{cases}
	a(k)\,e^{ikx}+b_+(k)\,e^{-ikx}\quad &,x<-(\ell+\delta)\\
	c_1(k)\,e^{i\tilde kx}+c_2(k)\,e^{-i\tilde kx}\quad &,-(\ell+\delta)\leq x<-\ell\\
	a_1(k)\cos(kx)+i\,a_2(k)\sin(kx) &,|x|\leq\ell\\
	c_3(k)\,e^{i\tilde kx}+c_4(k)\,e^{-i\tilde kx}\quad &,\ell<x\leq\ell+\delta\\
	e^{ikx} &,x>\ell+\delta,
  \end{cases}
\end{align}
where
\begin{align}
  a(k)&=a_1(k)a_2(k),\quad A(k)=\cos\tilde k\delta-\frac{i}{2}\bigg(\frac{k}{\tilde k}+\frac{\tilde k}{k}\bigg)\sin\tilde k\delta,\\
  a_{1/2}(k)&=e^{ik\delta}\bigg(A(k)\mp\frac{i}{2}e^{ik2\ell}\bigg(\frac{k}{\tilde k}-\frac{\tilde k}{k}\bigg)\sin\tilde k\delta\bigg),\\
%  a_2(k)&=e^{ik\delta}\bigg(A(k)+\frac{i}{2}e^{ik2\ell}\bigg(\frac{k}{\tilde k}-\frac{\tilde k}{k}\bigg)\sin\tilde k\delta\bigg),\\
  b_+(k)&=-\frac{i}{2}\bigg(\frac{k}{\tilde k}-\frac{\tilde k}{k}\bigg)\sin\tilde k\delta\big(A(k)e^{-ik2\ell}+\overline A(k)e^{ik2\ell}\big),\\
  c_{1/2}(k)&=e^{\pm i\tilde k\ell-ik(\ell-\delta)}\frac{1}{2}\bigg(1\pm\frac{k}{\tilde k}\bigg)A(k)-\frac{i}{4}e^{\pm i\tilde k\ell+ik(3\ell+\delta)}\bigg(1\mp\frac{k}{\tilde k}\bigg)\bigg(\frac{k}{\tilde k}-\frac{\tilde k}{k}\bigg)\sin\tilde k\delta,\\
%  c_2(k)&=e^{-i\tilde k\ell-ik(\ell-\delta)}\frac{1}{2}\bigg(1-\frac{k}{\tilde k}\bigg)A(k)-\frac{i}{4}e^{-i\tilde k\ell+ik(3\ell+\delta)}\bigg(1+\frac{k}{\tilde k}\bigg)\bigg(\frac{k}{\tilde k}-\frac{\tilde k}{k}\bigg)\sin\tilde k\delta,\\
  c_{3/4}(k)&=\frac{1}{2}\bigg(1\pm\frac{k}{\tilde k}\bigg)e^{i(k\mp\tilde k)(\ell+\delta)}\quad\text{and}\quad\tilde k=\sqrt{k^2-\lambda}.
\end{align}
%  c_4(k)&=\frac{1}{2}\bigg(1-\frac{k}{\tilde k}\bigg)e^{i(k+\tilde k)(\ell+\delta)},\\
\end{appendix}

%\bibliographystyle{plainnat}
%\bibliography{cit_resonances}
%merlin.mbs 2010-03-15 4.21a (PWD, AO, DPC)
\end{document}